# Resonant RF–Wakefield Coupling for Radiation-Reaction Control of 3D Betatron Dynamics in Hybrid Laser–Plasma Accelerators

A. A. Molavi Choobini[1], and M. Shahmansouri[1*]

[1]Department of Atomic and Molecular Physics, Faculty of Physics, Alzahra University, Tehran, Iran

**Abstract:**
  Hybrid laser–plasma–radiofrequency (RF) acceleration architectures signify a promising advancement in addressing the stability challenges associated with traditional laser wakefield accelerators. A thorough theoretical and numerical analysis of the three-dimensional dynamics of ultra-relativistic electron bunches in these hybrid systems is presented, clearly explaining how transverse beam stability, betatron oscillation polarisation, and radiative cooling work.  By combining analytical models of spatiotemporal plasma wakefield modulation and phase-dependent RF-driven oscillations with fully self-consistent 3D particle-in-cell (PIC) simulations, incorporating classical radiation reaction (RR) via the Landau–Lifshitz model (with quantum parameter $\chi_e$) to account for synchrotron-like losses during betatron oscillations. The findings indicate that the external RF fields operate as a tunable lattice, allowing for exact adjustment of amplitude, frequency, and carrier-envelope phase, which facilitates deterministic regulation of transverse focussing gradients and betatron amplitudes.  A regime of resonant alignment between RF fields and natural betatron frequencies is established; this resonance enhances controlled transverse excursions while concurrently diminishing parasitic oscillations via increased radiative damping, resulting in substantial emittance reduction and the alleviation of synchrotron-like energy losses.  Also, the detailed stability maps and 3D force landscapes show that the $\gamma$-factor growth rates change over time depending on the interaction between longitudinal field gradients and initial injection conditions.  The paper's results give a clear picture of the nonlinear, resonant, and damping events that happen in hybrid accelerators. They also make it possible to get ultra-stable, high-quality electron beams with the right polarisation states.

## I. Introduction

Plasma-based acceleration, encompassing laser wakefield accelerators (LWFAs) and plasma wakefield accelerators (PWFAs), has emerged as a transformative paradigm in high-gradient particle acceleration over the past two decades [1–3]. In these schemes, accelerating gradients on the order of 10–100 GV/m can be sustained, exceeding those achievable in conventional radiofrequency (RF) accelerators by several orders of magnitude. Such extreme gradients arise from the collective response of plasma electrons to an ultra-intense laser pulse or a relativistic charged-particle beam, which drives nonlinear plasma waves characterized by strong longitudinal electric fields. These wakefields enable the acceleration of electrons to GeV-scale energies within centimeter-scale distances, thereby opening pathways toward compact radiation sources, advanced light-generation platforms, and next-generation accelerator technologies for high-energy physics and medical applications. In addition to providing strong longitudinal acceleration, plasma wakefields inherently generate transverse focusing forces. These forces originate from the combined action of radial electric fields and azimuthal magnetic fields in the wake structure [4,5]. In the nonlinear blowout regime, the ion cavity formed behind the driver pulse creates an approximately linear focusing potential for relativistic electrons, leading to transverse oscillatory motion commonly referred to as betatron oscillations [6–10]. The characteristic betatron frequency in this regime can be expressed as $\omega_\beta = \kappa \omega_p / \sqrt{2\gamma}$, where $\omega_p$ stands for plasma frequency, $\kappa$ is a dimensionless parameter representing the focusing strength that can be derived from the wakefield structure, and $\gamma$ is the relativistic Lorentz factor of the electron. Betatron oscillations play a fundamental role in determining beam emittance, divergence, phase-space evolution, and radiation emission properties [11]. The transverse motion of relativistic electrons within the ion channel gives rise to synchrotron-like betatron radiation, which has been extensively explored as a compact source of ultrashort X-ray pulses. Beyond radiation generation, betatron dynamics strongly influence beam stability, transverse matching, and long-term energy evolution. Despite the remarkable progress achieved in plasma-based acceleration, several limitations continue to constrain beam performance and controllability [12–14]. The energy spread of accelerated electron bunches remains substantial due to longitudinal phase slippage, wakefield evolution, and injection dynamics. Transverse emittance growth can arise from mismatched injection conditions, evolving focusing gradients, and nonlinear transverse forces. Large-amplitude betatron oscillations may enhance radiative energy losses and amplify sensitivity to collective effects such as beam–plasma instabilities and space-charge interactions. These effects collectively degrade beam quality and complicate efforts to achieve precise phase-space control. As a result, there is strong motivation to explore mechanisms capable of externally regulating transverse motion, phase synchronization, and energy exchange within plasma accelerators.

Over the past decade, extensive theoretical, numerical, and experimental investigations have examined betatron dynamics across a wide range of accelerator regimes. In LWFA and PWFA regimes, numerous investigations have examined the intrinsic transverse dynamics arising from the nonlinear wake structure. Studies focusing on injection physics have shown how initial phase-space conditions strongly determine subsequent betatron amplitude and emittance evolution [15–18]. Plasma-density tailoring and longitudinal density ramps have been proposed to mitigate phase slippage and control oscillation amplitude [19, 20]. These works clarified the dependence of betatron frequency and focusing strength on plasma density and relativistic factor, yet the transverse restoring forces were treated as purely wakefield-determined quantities, without external modulation. Parallel research efforts have explored radiation emission associated with betatron motion. Three-dimensional PIC simulations in both underdense and near-critical-density plasmas demonstrated enhanced transverse oscillations leading to broadband X-ray radiation [17, 21-23]. Analytical and numerical approaches based on Liénard–Wiechert potentials enabled detailed reconstruction of angular–spectral radiation properties from simulated trajectories [24]. Experimental studies further examined coherence properties and proposed methods for emittance reconstruction through radiation diagnostics by Alessandro Curcio and colleagues [25]. While these investigations advanced understanding of emission mechanisms, the betatron motion itself remained governed solely by plasma self-fields. In addition, researchers addressed controlled beam manipulation in plasma accelerators. Strategies such as periodic wake loading and staged acceleration were proposed to enhance luminosity and stabilize beam transport [26]. Post-acceleration beam conditioning using magnetic chicanes and conventional RF elements was also investigated to manipulate longitudinal phase space after plasma acceleration [27]. In addition, studies in direct laser acceleration (DLA) demonstrated that coupling between the laser field and betatron oscillations can enhance energy gain under specific resonance conditions [17, 20]. However, in DLA the driving electromagnetic field is intrinsically tied to the laser pulse that excites the plasma, limiting independent tunability of frequency, phase, and polarization.

Despite these advances, a clear separation persists between plasma-driven transverse focusing and externally controllable electromagnetic modulation. In conventional RF accelerators, phase stability and energy synchronization are achieved through precisely engineered RF cavities, enabling deterministic control over longitudinal dynamics. In contrast, plasma wakefield accelerators rely predominantly on self-consistent plasma fields, whose structure evolves dynamically and offers limited external adjustability. To date, however, no systematic framework has been established in which an independently controlled RF electromagnetic field is explicitly incorporated into the transverse restoring forces of a nonlinear plasma wake to directly regulate betatron dynamics. In this work, a hybrid laser–plasma–RF acceleration framework is introduced and systematically analyzed in which externally applied RF electromagnetic fields are superimposed onto a nonlinear plasma wake. Although plasma accelerators sustain gradients orders of magnitude higher than conventional RF structures, the present hybrid scheme does not require RF components to operate at plasma-level fields. The plasma wake remains the dominant acceleration mechanism, while the externally applied RF field functions as a moderate-amplitude, phase-controlled modulation of transverse focusing and betatron motion. Such fields can be introduced through transverse injection geometries or dielectric/photonic coupling structures that remain spatially separated from the highest-gradient plasma region, without requiring detailed modeling of specific RF cavity architectures. Hybrid RF–plasma concepts have been discussed in staged accelerator designs and THz-assisted modulation schemes, demonstrating that controlled electromagnetic fields can be integrated with plasma accelerators without exceeding material breakdown limits. By embedding RF contributions directly into the effective restoring forces, the model extends the conventional betatron evolution equations and modifies the effective focusing strength, phase evolution, and energy exchange mechanisms experienced by the electrons. The analytical framework developed here allows for arbitrary RF phase, frequency, and polarization, thereby enabling systematic investigation of resonant coupling conditions between the imposed RF frequency and the intrinsic betatron frequency. Under appropriate synchronization conditions, the RF field can induce controlled modulation of betatron amplitude and phase, alter the polarization state of transverse motion, and modify the cycle-averaged energy evolution relative to purely plasma-driven acceleration. In particular, the interplay between plasma focusing and RF forcing can lead to transitions between linear, elliptical, and near-circular transverse trajectories, revealing new regimes of controllable oscillatory dynamics. To validate and quantify these analytical predictions, fully self-consistent three-dimensional PIC simulations are performed. The numerical framework simultaneously incorporates plasma wake excitation by the driving laser pulse, externally applied RF electromagnetic fields, and radiation-reaction effects influencing ultra-relativistic electron motion. This unified analytical–computational approach enables construction of stability maps and effective force landscapes that identify parameter regimes where RF–plasma coupling enhances, suppresses, or stabilizes transverse oscillations. The simulations further clarify how phase synchronization influences long-term energy evolution, transverse confinement, and modulation efficiency. The results demonstrate that RF-driven modulation provides a tunable mechanism for regulating betatron amplitude, phase stability, and energy exchange in ultra-relativistic regimes. By explicitly bridging plasma wakefield physics with concepts of externally controlled RF modulation, the present work establishes a unified framework that connects two traditionally distinct accelerator communities within a single dynamical description. The paper is organized as follows: In section II, the theory of the mechanism on Betatron dynamics is presented. Discussion on radiation reaction in hybrid RF–laser plasma accelerators are given in section III. Conclusions are drawn in section IV.

## II. Theoretical Model

A model of hybrid laser–plasma–RF accelerators, enabling the modulation of beam quality, radiation losses, and betatron polarization to investigate a theoretical model of betatron oscillations and radiation reaction in laser-driven plasma accelerators, is constructed. The resulting formalism supports the development of ultra-relativistic electron sources with precise control over transverse dynamics. For that, consider a relativistic electron ($\gamma \gg 1$) that trapped in a plasma with the longitudinal independent variables $\xi = k_p(z - \beta_{ph}ct)$ (co-moving longitudinal coordinate and dimensionless). The variables $z$ and $t$ denote the longitudinal coordinate and time, respectively; $c$ is the speed of light, where $\beta_{ph}$ is the phase velocity and normalized to $c$. Then, the normalized RF electric field as a spatiotemporal transverse wave given by:

$$\vec{E}_{RF}(\xi,\tau) = E_{0x}^{RF} Cos\varphi_{RF}\hat{e}_x + E_{0y}^{RF} Cos(\varphi_{RF} + \theta)\hat{e}_y \qquad (1)$$

where the $E_{0x}^{RF}$ and $E_{0y}^{RF}$ are the amplitudes of the RF electric field in the $x$ and $y$ directions, respectively. $\varphi_{RF} = \hat{k}_{RF}\xi - \hat{\omega}_{RF}\tau$, $\tau = \omega_p t$, $\hat{\omega}_{RF} = \omega_{RF}/\omega_p$, $\hat{k}_{RF} = k_{RF}/k_p$, $k_{RF}$ is the RF wavenumber, $\omega_{RF}$ is the angular frequency, and $\theta$ is the phase difference between components. The phase $\theta$ allows for elliptical polarization, enabling control over the beam's transverse dynamics and the polarization of emitted radiation. It should emphasize that Eq. (1) does not represent a full RF cavity eigenmode, such as a standing-wave TM structure commonly used in conventional RF accelerators. Instead, the adopted field configuration corresponds to a reduced traveling-wave electromagnetic field employed to model an externally applied RF modulation in the hybrid plasma–RF framework considered here. Such traveling-wave representations are widely used as analytical approximations of disk-loaded or dielectric-loaded accelerating structures when the primary objective is to investigate phase-dependent beam–field interaction rather than detailed cavity eigenmode formation [28, 29]. In the present work, the RF field acts as a controlled transverse perturbation superimposed on the nonlinear plasma wakefield rather than serving as the primary accelerating structure. This approximation enables transparent analytical treatment of phase synchronization and resonant betatron modulation without introducing full cavity-mode expansions. The magnetic component associated with the RF field is obtained self-consistently from Maxwell's equations. Accordingly, the full Lorentz force is retained in the analytical formulation. In the ultra-relativistic limit, the transverse force contribution reduces to an effective enhancement of the electric-field term, which is consistently incorporated into the normalization of the RF modulation amplitude used throughout the analysis. Therefore, neglecting inter-particle interactions, the electric fields generated by the plasma wakefield are expressed as:

$$E_\parallel = E_\parallel|_{\langle\xi\rangle} + \frac{\partial E_\parallel}{\partial \xi}|_{\langle\xi\rangle}\xi_1 \qquad (2a)$$

$$\vec{E}_\perp = \vec{E}_{RF}(z,t) + \Omega_\beta^2(\xi,\tau,\gamma)\left(1 - \frac{\partial E_\parallel}{\partial \xi}|_{\langle\xi\rangle}\right)\vec{r} \qquad (2b)$$

$$\vec{B}_\perp = \vec{B}_{RF}(z,t) + \vec{B}_\theta \qquad (2c)$$

where $E_\parallel|_{\langle\xi\rangle}$ is the static longitudinal electric field at the mean wake position $\langle\xi\rangle$, $\frac{\partial E_\parallel}{\partial \xi}|_{\langle\xi\rangle}$ is the field gradient, $\xi_1 = \xi - \langle\xi\rangle$ is the deviation from the mean position, $\Omega_\beta = \omega_p/\sqrt{2}$ is the plasma focusing constant with the plasma frequency ($\omega_p$). $\vec{B}_\perp$ is the transverse magnetic field in the hybrid system, $\vec{B}_\theta = -\Omega_\beta^2(\xi,\tau,\gamma)\frac{\partial E_\parallel}{\partial \xi}|_{\langle\xi\rangle} r\hat{e}_\theta$ is the plasma-focusing magnetic field, $\vec{B}_{RF}(z,t)$ is the RF magnetic field, and $\gamma$ is the Lorentz factor. The longitudinal field accelerates the electron along $z$, while the transverse fields provide focusing, with the RF field adding an oscillatory component. After interaction, the focusing strength $\Omega_\beta(\xi,\tau,\gamma)$, which depends on the net transverse electric field, is modulated by the RF field and should carry the same spatial and temporal phase as follows:

$$\Omega_\beta(\xi,\tau,\gamma) = \frac{\tilde{\kappa}_0^2}{\gamma}[1 + \delta_\kappa Cos\varphi_{RF}] \tag{3}$$

where $\tilde{\kappa}_0$ is the baseline focusing strength from the plasma wakefield, and $\delta_\kappa \ll 1$ is the RF-induced modulation amplitude (dimensionless). The focusing force in a plasma accelerator arises from transverse electric and magnetic fields. These modified focusing show up as a modulation in $\Omega_\beta(\xi,\tau,\gamma)$, leading to betatron frequency modulation and, potentially, RF-induced parametric effects in the transverse dynamics. Therefore, due to this, the force can be expressed as:

$$F_x = -m_e\Omega_\beta^2\left[1 - \frac{\partial E_\parallel}{\partial \xi}|_{\langle\xi\rangle}\left(1 - \frac{\omega_p\Omega_\beta}{\sqrt{\gamma}}\right)\right]x + eE_{0x}^{RF}Cos\varphi_{RF} + F_x^R \tag{4a}$$

$$F_y = -m_e\Omega_\beta^2\left[1 - \frac{\partial E_\parallel}{\partial \xi}|_{\langle\xi\rangle}\left(1 - \frac{\omega_p\Omega_\beta}{\sqrt{\gamma}}\right)\right]y + eE_{0y}^{RF}Cos(\varphi_{RF} + \theta) + F_y^R \tag{4b}$$

$$F_z = -E_\parallel|_{\langle\xi\rangle} - \frac{\partial E_\parallel}{\partial \xi}|_{\langle\xi\rangle}\xi_1 + \Omega_\beta^2\frac{\partial E_\parallel}{\partial \xi}|_{\langle\xi\rangle}(x\dot{x} - y\dot{y}) + F_z^R \tag{4c}$$

here $\vec{F}^R$ is the radiation reaction force ($F_{rr}$), a dot on the top means taking the time derivative, the subscript 1 means the betatron oscillation (BO) term, and the term $\omega_p\Omega_\beta/\sqrt{\gamma}$ accounts for relativistic effects. The first, second, and third terms on the right-hand side of Eq. (4) correspond respectively to plasma focusing, RF driving, and the RR correction described using the Landau–Lifshitz formulation. The plasma term provides the approximately linear focusing response of the ion cavity, the RF term introduces an externally controlled transverse modulation, and the remaining contribution accounts for radiative damping. The transverse forces include a restoring force from plasma focusing, an RF-driven oscillatory term, and radiation reaction, while the longitudinal force drives acceleration and couples' transverse and longitudinal motion.

To describe the energy evolution, the rate of change of the Lorentz factor under the influence of the RF field, plasma wake, and radiation losses is considered. In the limit $r\gamma\rho_e/2a \ll 1$, where $\rho_e = k_p r_e$ is classical radius parameter, $r_e$ is the classical electron radius, and $a$ is the normalized vector potential of the laser, the modified energy equation becomes:

$$\dot{\gamma} = \frac{e}{m_e c}\vec{E}_{RF}\cdot\vec{v}_\perp$$

$$-E_\parallel|_{\langle\xi\rangle}\left(\frac{\omega_p\Omega_\beta}{\sqrt{\gamma}}\right) - \frac{2}{3}\rho_e\gamma^2\Omega_\beta^4(x^2+y^2) + \left[\frac{\partial E_\parallel}{\partial \xi}|_{\langle\xi\rangle}\left(\Omega_\beta^2 + \frac{1}{4}\left(\frac{\omega_p\Omega_\beta}{\sqrt{\gamma}}\right)\right) - \Omega_\beta^2\right](x\dot{x}+y\dot{y}) \tag{5}$$

where $\vec{v}_\perp = \dot{x}\hat{e}_x + \dot{y}\hat{e}_y$. The first term on the right-hand side of Eq. (5) represents RF energy gain, the second term corresponds to plasma-wake acceleration, the third term represents radiative damping, and the fourth term denotes transverse–longitudinal coupling. By inserting the Lorentz force [Eqs. (4a–4c)] into the momentum equation, separating transverse and longitudinal components, usage of the energy evolution equation [Eq. (5)], and with assuming slow variation of A compared to the betatron period and averaging over the fast oscillations, the momentum evolution follows from the forces:

$$\dot{P}_x = eE_{0x}^{RF} Cos\varphi_{RF} + \Omega_\beta^2 x \left[\frac{1}{2}\frac{\partial E_\parallel}{\partial \xi}|_{\langle\xi\rangle}(\langle\gamma\rangle^{-2} + \dot{x}^2 + \dot{y}^2) - 1\right] - \frac{2}{3}\rho_e\gamma^2\Omega_\beta^4(x^2 + y^2)\dot{x} \quad (6a)$$

$$\dot{P}_y = eE_{0y}^{RF} Cos(\varphi_{RF} + \theta) + \Omega_\beta^2 y \left[\frac{1}{2}\frac{\partial E_\parallel}{\partial \xi}|_{\langle\xi\rangle}(\langle\gamma\rangle^{-2} + \dot{x}^2 + \dot{y}^2) - 1\right] - \frac{2}{3}\rho_e\gamma^2\Omega_\beta^4(x^2 + y^2)\dot{y} \quad (6b)$$

$$\dot{P}_z = -E_\parallel|_{\langle\xi\rangle} + \frac{\partial E_\parallel}{\partial \xi}|_{\langle\xi\rangle}\left(\Omega_\beta^2 + \frac{1}{4}\right)(x\dot{x} + y\dot{y}) - \frac{2}{3}\rho_e\gamma^2\Omega_\beta^4(x^2 + y^2) \quad (6c)$$

where $\langle\gamma\rangle$ is the average Lorentz factor. These equations describe the interplay of RF-driven motion, plasma focusing, and radiation reaction. To simplify the transverse dynamics, the complex amplitudes is defined as follows:

$$\Phi = \left(x + \frac{\sqrt{\gamma}}{i\Omega_\beta}\dot{x}\right)e^{-i\Theta} \quad (7a)$$

$$\Psi = \left(y + \frac{\sqrt{\gamma}}{i\Omega_\beta}\dot{y}\right)e^{-i\Theta} \quad (7b)$$

$$\Theta = \int \frac{\Omega_\beta}{\sqrt{\gamma}} dt \quad (7c)$$

where $\Phi, \Psi$ represent the transverse oscillations in the $x$ and $y$ directions, respectively, and $\Theta$ is the betatron phase. Taking the time derivative of $\Phi$, one can obtain:

$$\dot{\Phi} = \frac{ie}{2\kappa\sqrt{\gamma}}\left[E_{0x}^{RF} Cos\varphi_{RF} + iE_{0y}^{RF} Cos(\varphi_{RF} + \theta)\right]e^{-i\Theta} + \frac{1}{4}E_{0\parallel}\frac{\omega_p\Omega_\beta}{\sqrt{\gamma}}\langle\gamma\rangle^{-1}\Phi - \frac{1}{24}k_{pre}\Omega_\beta^4\langle\gamma\rangle[|\Phi|^2\Phi + 2|\Psi|^2\Phi - \Psi^2\Phi] + \frac{i}{64}\frac{\omega_p\Omega_\beta^2}{\sqrt{\gamma}}\langle\gamma\rangle^{-3/2}[|\Phi|^2\Phi + |\Psi|^2\Phi] - \frac{i}{16}\Omega_\beta^3\langle\gamma\rangle^{-3/2}\left[\left(|\Phi|^2 + 2\frac{\partial E_\parallel}{\partial \xi}|_{\langle\xi\rangle}|\Psi|^2\right)\Phi - \left(2\frac{\partial E_\parallel}{\partial \xi}|_{\langle\xi\rangle} - 1\right)\Psi^2\Phi\right] - \frac{i}{4}\frac{\partial E_\parallel}{\partial \xi}|_{\langle\xi\rangle}\Omega_\beta\langle\gamma\rangle^{-5/2}\Phi \quad (8)$$

where $E_{0\parallel}$ is the longitudinal field amplitude. The Eq. 8, derived by substituting the force and energy equations into the time derivative of $\Phi$, captures RF modulation, plasma effects, and nonlinear radiation reaction. Then, The averaged energy and longitudinal dynamics are:

$$\langle\dot{\gamma}\rangle = e\vec{E}_{RF}\cdot\vec{v}_\perp - E_{0\parallel}\frac{\omega_p\Omega_\beta}{\sqrt{\gamma}} - \frac{1}{3}k_{pre}\Omega_\beta^4\langle\gamma\rangle^2[|\Phi|^2 + |\Psi|^2] \quad (9)$$

$$\langle\dot{\xi}\rangle = \frac{1}{2(\gamma_w^2 - \langle\gamma\rangle^2)} - \frac{\Omega_\beta^2}{4\langle\gamma\rangle}[|\Phi|^2 + |\Psi|^2] \quad (10)$$

where $\gamma_w = 1/\sqrt{1 - \beta_{ph}^2}$ is the Lorentz factor of the wake. Equation (9), obtained by averaging over betatron oscillations, link energy evolution to transverse dynamics and radiation losses. The effective longitudinal force includes RF-induced phase-dependent contributions, allowing modulation of energy gain at fixed plasma density. To account for polarization, the expressions of $\Phi = |\Phi|e^{i\varphi_x}$, and $\Psi = |\Psi|e^{i\varphi_y}$ is defined, where $\Delta\varphi = \varphi_y - \varphi_x$. Therefore, the amplitude and phase dynamics are:

$$\frac{\partial |\Phi|}{\partial t} = \{\frac{e}{2\Omega_\beta\sqrt{\gamma}}\left[E_{0y}^{RF} Cos\varphi_{RF} Cos(\phi - \varphi_x) + E_{0y}^{RF} Cos(\varphi_{RF} + \theta) Cos(\phi - \varphi_y)\right]$$
$$+ \frac{1}{4\langle\gamma\rangle}E_{0\parallel}\frac{\omega_p\Omega_\beta}{\sqrt{\gamma}}|\Phi| - \frac{1}{24}k_{pre}\Omega_\beta^4\langle\gamma\rangle[|\Phi|^3 + |\Psi|^2|\Phi|(2Cos(2\Delta\varphi))]$$
$$- \frac{\Omega_\beta}{16}\left(\frac{1}{4}\frac{\partial E_\parallel}{\partial \xi}|_{\langle\xi\rangle}\frac{\omega_p\Omega_\beta}{\sqrt{\gamma}} - \Omega_\beta^2\left(1 - 2\frac{\partial E_\parallel}{\partial \xi}|_{\langle\xi\rangle}\right)\right)\langle\gamma\rangle^{-3/2}|\Psi|^2\Phi Sin(2\Delta\varphi)\}e^{-i\Theta} \quad (11)$$

where $\phi$ is a reference phase. This form naturally accounts for elliptical polarization and phase delay between $x$ and $y$ components, as follows:

$$\frac{\partial \varphi_x}{\partial t} = \frac{1}{24}\rho_e \Omega_\beta^4 \langle\gamma\rangle |\Psi|^2 Sin(2\Delta\varphi) + \frac{1}{64}\frac{\partial E_\parallel}{\partial \xi}|_{\langle\xi\rangle}\frac{\omega_p \Omega_\beta^2}{\sqrt{\gamma}}\langle\gamma\rangle^{-3/2}[|\Phi|^2 + |\Psi|^2 Cos(2\Delta\varphi)] -$$
$$\frac{1}{16}\Omega_\beta^3 \langle\gamma\rangle^{-\frac{3}{2}}\left[|\Phi|^2 + 2\frac{\partial E_\parallel}{\partial \xi}|_{\langle\xi\rangle}|\Psi|^2 + \left(1 - 2\frac{\partial E_\parallel}{\partial \xi}|_{\langle\xi\rangle}\right)|\Psi|^2 Cos(2\Delta\varphi)\right] - \frac{\Omega_\beta}{4}\frac{\partial E_\parallel}{\partial \xi}|_{\langle\xi\rangle}\langle\gamma\rangle^{-5/2} \quad (12)$$

and

$$\frac{\partial \Delta\varphi}{\partial t} = -\frac{1}{24}\rho_e \Omega_\beta^4 \langle\gamma\rangle[|\Psi|^2 + |\Phi|^2]Sin(2\Delta\varphi)$$
$$+ \frac{\Omega_\beta}{8}\left[\frac{1}{4}\frac{\partial E_\parallel}{\partial \xi}|_{\langle\xi\rangle}\frac{\omega_p \Omega_\beta}{\sqrt{\gamma}} - \Omega_\beta^2\left(1 - 2\frac{\partial E_\parallel}{\partial \xi}|_{\langle\xi\rangle}\right)\right]\langle\gamma\rangle^{-3/2}[|\Psi|^2 - |\Phi|^2]Sin^2\Delta\varphi \quad (13)$$

These equations, derived by separating the real and imaginary parts of $\dot{\Phi}$ and $\dot{\Psi}$, describe the evolution of the oscillation amplitudes and phases, capturing elliptical polarization and phase delays critical for radiation properties.

## III. Results and Discussion

In this study, the feasibility of modulating beam quality, radiation losses, and betatron polarization in hybrid laser–plasma–RF accelerators through a comprehensive theoretical and numerical framework are demonstrated, highlighting its potential for next-generation ultra-relativistic electron sources with enhanced transverse control. To validate and extend the analytical results, fully self-consistent 3D PIC simulations including radiation reaction effects, which in the $\chi \ll 1$ regime reproduce classical Landau–Lifshitz damping. By enabling (i) arbitrary RF phase control in 3D, (ii) simultaneous treatment of RF modulation, plasma focusing, and nonlinear RR, and (iii) extraction of full 3D force landscapes and stability maps, combined analytical–numerical treatment has been done.

It should emphasize that the hybrid RF–plasma framework investigated here naturally supports two distinct operational regimes within a single, unified physical model. In the first regime, phase-synchronized RF modulation acts in a damping-dominated manner, suppressing large-amplitude betatron oscillations to improve transverse beam quality and reduce normalized emittance. In the second regime, resonant coupling between the RF field and the intrinsic betatron frequency leads to controlled amplification of transverse oscillations, which can be exploited to enhance betatron radiation emission. Both regimes are treated self-consistently within the same analytical formulation and PIC simulation framework, with their respective outcomes determined by the choice of RF frequency, phase, amplitude, and polarization.

The analytical model encompasses the spatiotemporal RF electric field (Eq. 1), plasma wakefield expressions (Eqs. 2a–c), RF modulated focusing strength (Eq. 3), transverse and longitudinal equations of motion including radiation reaction (Eqs. 4a–c), energy evolution (Eq. 5), momentum dynamics (Eqs. 6a–c), complex betatron amplitudes (Eqs. 7a–c), and phase-dependent amplitude equations (Eqs. 8–13). In the analytical formulation, radiation reaction is written explicitly in the deterministic Landau–Lifshitz form to provide physical transparency.

The particle-in-cell (PIC) simulations were performed using the EPOCH code (development branch based on version 4.20-devel, circa mid-2025; https://github.com/Warwick-Plasma/epoch) [30]. These simulations self-consistently incorporated radiation reaction via the built-in module operating in the classical limit ($\chi \approx 10^{-7} \ll 1$, reproducing Landau-Lifshitz damping without stochastic emission). The configuration used a 3D Cartesian grid with 512 x 256 x 256 cells in the x (longitudinal, laser propagation), y, and z (transverse) directions, respectively, with uniform cell size $\Delta x = \Delta y = \Delta z = 0.05\ c/\omega_p \approx 0.084\ \mu m$ (for $n_e = 10^{18}\ cm^{-3}$, $\omega_p \approx 1.78 \times 10^{14}\ rad/s$, corresponding to a skin depth $c/\omega_p \approx 1.69\ \mu m$). This yields a physical domain size of $\approx 43\mu m$ (longitudinal) × 21.5μm × 21.5μm (transverse). A moving window with velocity $v_w = 0.999c$ was employed to track the interaction over propagation distances up to ~1mm ($t_{max} \approx 3.3ps$).

According to Fig.1, the driving laser was a Gaussian pulse with peak normalized vector potential $a_0 = 2$ (peak intensity $I \approx 8.6 \times 10^{18} W/cm^{-2}$ at $\lambda = 800\ nm$), pulse duration $\tau = 30fs$ (FWHM), with a Gaussian temporal profile $exp\left[-\frac{t^2}{(\tau/\sqrt{2Ln2})^2}\right]$, and a focal spot size $w_0 = 10\mu m$ (FWHM, Gaussian transverse profile), focused at the plasma entrance (x = 0). The plasma was a uniform slab ($n_e = 10^{18}\ cm^{-3}$, and immobile ions) starting at x = 20 μm with open transverse boundaries and absorbing longitudinal boundaries. The electron bunch was self-injected from background plasma electrons in the blowout regime (no external injection), with post-injection parameters at $\tau \approx 100$ given by an average Lorentz factor $\gamma_0 \approx 100$ (~50 MeV), relative energy spread $\Delta\gamma/\gamma \approx 15 - 25\%$, total charge Q ≈ 10 pC, rms length $\sigma_x \approx 1\mu m$, transverse rms sizes $\sigma_y = \sigma_z \approx 0.5\mu m$, normalized emittance $\varepsilon_n \approx 1 - 5\mu m$. The RF modulation was injected as an external transverse electromagnetic plane wave via EPOCH's laser boundary condition (amplitudes $E_{RF} = 10 - 100\ MV/m$ corresponding to normalized modulation strengths $\alpha \approx 0.01 - 0.1$, frequency $\omega_{RF}$ matched to $\omega_\beta$ in THz range, tunable phase $\phi$ and polarization $\delta$), synchronized with the laser driver for resonant coupling. Macroparticles per cell: 8 for electrons and 8 for ions (16 total). Outputs in SDF format were post-processed for averaged quantities (e.g., $\langle\gamma\rangle$ from Eq. 9) and ensemble trajectories (~$10^4$ particles tracked). The quantum parameter $\chi_e = 2\gamma E_\perp \hbar\omega_p/3m_e^2 c^2 \approx 10^{-7}$ is small, making quantum corrections negligible over t_max≈ 3ps, but classical RR provides measurable damping [31, 32]. To address feasibility, RF fields penetrate the plasma via THz frequencies matched to $\omega_p$, avoiding strong screening for $\omega_{RF} < \omega_p$. Synchronization is achieved via phase-locking the RF to the laser driver, and coupling geometry uses transverse injection or photonic structures for efficient field-plasma overlap [33, 34].

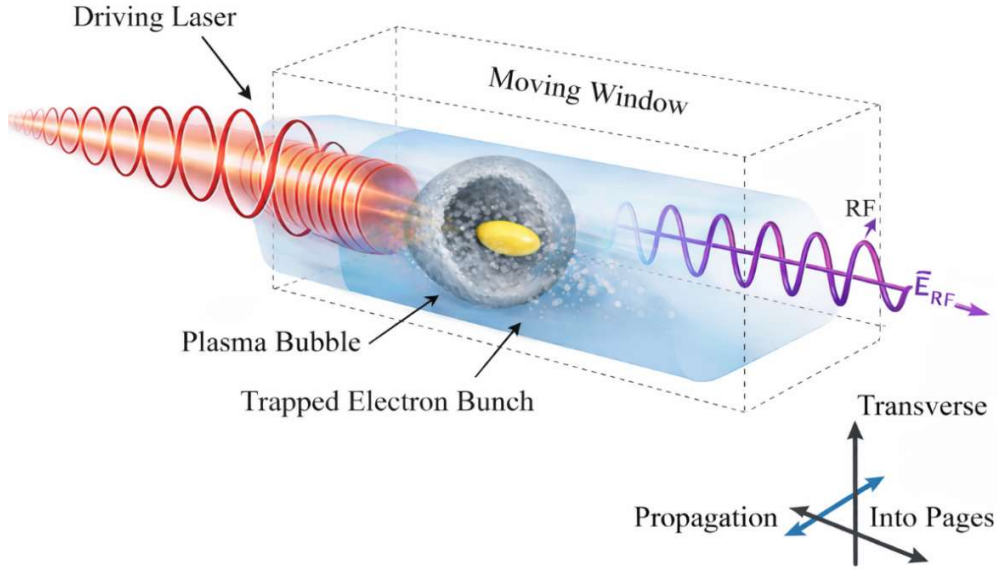

**Fig. 1.** Schematic of the hybrid laser–plasma–RF accelerator simulation geometry. The driving laser pulse propagates along x, exciting a blowout wakefield in the uniform plasma. The trapped electron bunch is self-injected in the first bucket. Transverse RF modulation is applied for resonant betatron control. The moving window tracks the interaction at $v_w \approx c$. Axes: x (propagation), y (transverse vertical), z (into page).

No externally initialized electron bunch was used in simulations. All plasma electrons were initially loaded at rest. As the laser pulse propagates through the uniform plasma, it self-consistently drives a nonlinear wakefield in the blowout regime, leading to self-injection of background plasma electrons. Injection occurs at the rear of the first wake bucket when a fraction of sheath electrons becomes trapped in the accelerating phase of the wake. No density down-ramp or auxiliary injection mechanism was employed. The total electromagnetic fields experienced by the particles in the PIC simulations ,therefore, consist solely of the self-generated plasma fields and the externally injected RF field, without any imposed analytical wake structure. The electromagnetic fields had evolved using EPOCH's standard Yee finite-difference time-domain (FDTD) Maxwell solver. To mitigate potential numerical Cherenkov instability in the long propagation regime with relativistic electrons, the Courant–Friedrichs–Lewy factor was set to 0.95–0.98, standard binomial current smoothing (1–2 passes) was applied, and the longitudinal resolution exceeded 9 cells per laser wavelength. Throughout the ~3 ps propagation, no signatures of numerical Cherenkov instability, such as spurious high-frequency radiation, artificial emittance growth, or non-physical phase-space heating, were observed. The figures are primarily derived from numerical integration of the analytical models (Equations 1–13) using fourth-order Runge–Kutta solvers (implemented in MATLAB/Python), with parameter ranges chosen to match the PIC setup for direct validation. Integrations cover normalized time $\tau = \omega_p t$ up to 100–1000 (~$10^2$–$10^3$ betatron periods) to capture resonances and damping. Trends and resonance conditions were cross-validated against post-processed 3D PIC data from EPOCH, including ensemble-averaged trajectories (~$10^4$ macroparticles), field extractions, and stability metrics.

In Figure 2, one sees how the transverse dynamics and polarisation effects together in a hybrid laser-plasma-RF accelerator system. These plots, obtained from numerical integration of Equations (8)–(13) with parameters matched to the PIC setup and validated against ensemble-averaged PIC trajectories, illustrate how electron oscillations, focussing forces, and radiation properties all affects. Figure 2a shows that the normalised radius changes smoothly when the level of modulation changes. Due to changes the focussing strength by time, the cross-section of the electron beam changes as well. This kind of behaviour indicates how sensitive the system is to RF-induced modulations and how they directly affect betatron oscillations. As the modulation amplitude increases, the oscillatory motion becomes stronger, which influence beam quality and stability. Figure 2b presents a symmetric pattern of the polarization parameter as a function of phase difference. A distinct central peak is observed, which broadens with increasing RF strength. It implies that the elliptical polarization of the radiation emitted should be very carefully controlled by changing not only the phase but also the amplitude of the RF field - a crucial aspect for the control of radiation properties. The hourglass-like figure shows a resonance condition where polarization effects are the strongest, thus giving a way for the exact control of the beam's transverse dynamics. Figure 2c illustrates damped oscillatory trajectories of electrons in three-dimensional space. Here, the radius decreases with stronger RF fields and small phase variations, demonstrating the damping effect of higher frequencies on betatron oscillations. The distinct trajectories observed for different plasma and RF frequencies highlight the importance of frequency matching. This confirms the system's capability to sustain ultra-relativistic electron beams while maintaining precise control over their transverse motion. The resulting transverse dynamics is consistent with experimental observations, reinforcing the connection between plasma response dynamics [35 - 37].

The detailed exploration of the modulated focusing strength within the hybrid laser–plasma–RF accelerator system is provided in Figure 3. It derived from numerical solution of Equation (3) and validated by PIC field extractions and shedding light on how the transverse dynamics of electron beams are influenced by RF-induced perturbations and temporal evolution. In Figure 3a, the change of the focusing strength with the normalized longitudinal coordinate looks to be oscillatory and the oscillations represent the periodic increase and decrease. The size of these oscillations becomes larger as the RF modulation parameter increases; thus, it is suggested that higher modulation amplitudes lead to the larger variations of the focusing, which then cause stronger betatron oscillations that may even affect the stability of the beam. The different lines for various $\delta_k$ values signify a controllable tuning mechanism whereby the system can be set to a condition that the focusing strength is at its optimum for certain accelerator situations thus the RF fields playing the role of the fine-tuning agent in the plasma wakefield's focusing properties. Figure 3b extends this analysis into a three-dimensional representation, where the focusing strength exhibits a smooth, wave-like surface that peaks at intermediate values of both modulation amplitude and normalized time. This implies that there is a resonant interaction where the focussing is at its best under certain conditions. This is due to the RF field and plasma wake are in phase with each other. The gradual rise and fall across the surface suggest that the focussing force changes over time, thanks to the interaction of RF modulation and plasma frequency. This could be very important for keeping the beam coherent and reducing radiation losses during acceleration. Each trajectory corresponds to an independent PIC simulation initialized with a different plasma density or $\omega_p$, with particles tracked shortly after self-injection ($\gamma \approx 2-5$ at the onset of tracking).

The set figures of 4 imply how the plasma focussing affects and how stable it is in the hybrid laser-plasma-RF accelerator system. Figure 3 presents analytical evaluations and stability analysis from Equations (6–8), cross-validated with PIC growth rates. They also teach us a lot about how relativistic electron beams behave in different situations. Figure 4a shows that the plasma focussing coefficient drops quickly when the Lorentz factor goes up. The plasma frequency affects how quickly the rate of decrease is. This indicates that until higher frequencies start to drop off, at lower relativistic energies, it makes the focussing strength stronger. This shows that plasma effects don't work as well when electrons are close to ultra-relativistic limits. This trend indicates a potential limit to the effectiveness of plasma focusing on high energies, which could impact the beam's transverse confinement. Figure 4b delves into this by indicating how the focusing coefficient changes for different longitudinal field gradients that ranges of Lorentz factors are broader for the focusing strength that gradients steeper sustain. As a result, the spatial structure of the wakefield becomes the main lever to ensure beam stability which, in turn, can be used for performance optimization of acceleration. In Figure 4c, ones can see a stability landscape where the growth rate of instabilities is highest at middle values of the normalised longitudinal coordinate and modulation amplitude. This shows areas where instability could happen and affect beam coherence. The smooth gradient shows that the transition is happening slowly, but it can be sped up by carefully changing the parameters. Figure 4d supports this by showing that stability is present across RF phase and modulation amplitude. As the phase angle gets bigger, the stability parameter gets bigger too. It has a complicated relationship with modulation. In the RF field, phase alignment and amplitude modulation are very important for keeping the beam stable and stopping it from moving around. This gives us a way to make the system more reliable when it needs a lot of energy. The results of modulated focussing strength, along with theoretical predictions and PIC simulations, match what was found in experiments [38, 39].

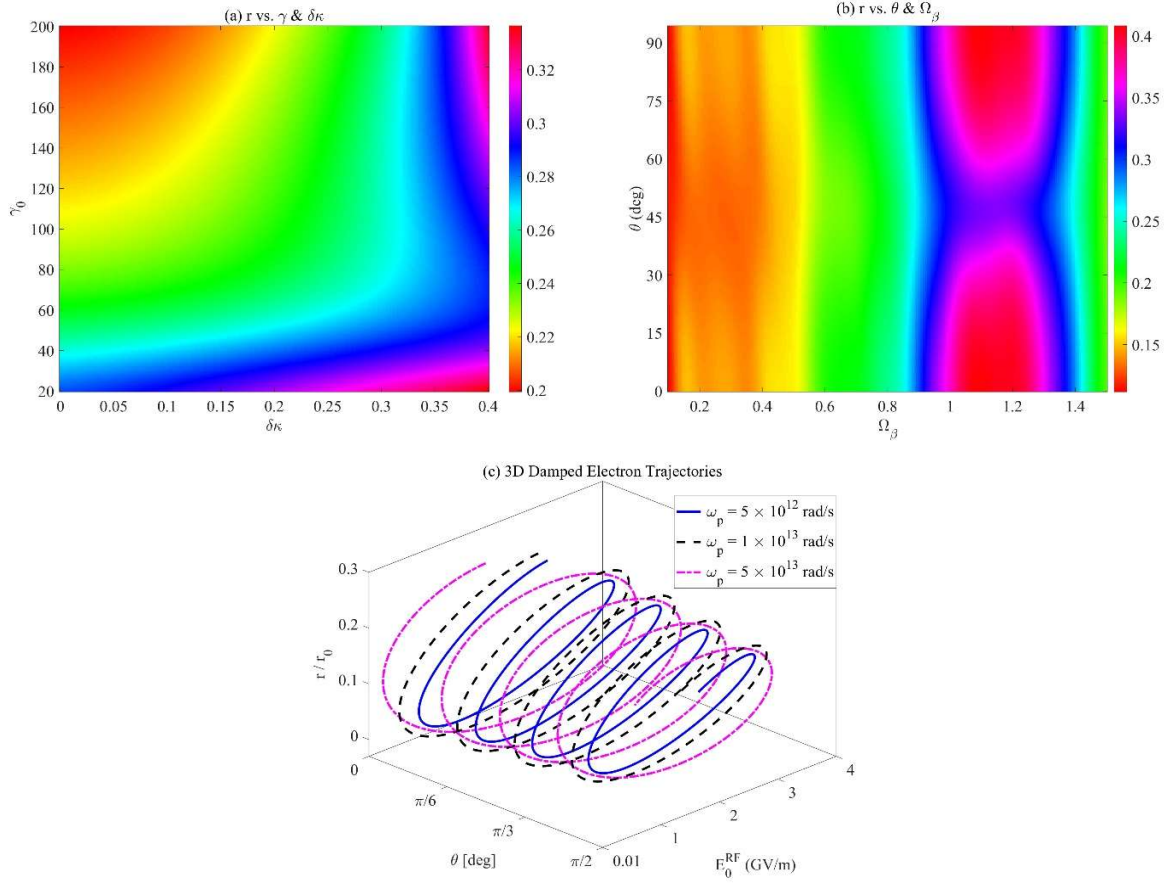

**Fig. 2.** (a) color map of normalized radius trajectories of electrons as a function of modulation amplitude and initial relativistic factor, (b) color map of normalized radius trajectories of electrons as a function of phase difference and strength parameter, (c) 3D plot of normalized radius trajectories of electrons under varying RF electric field and phase, with different plasma and RF frequencies. The three trajectories are representative electrons selected from the self-injected population in the first wake bucket, chosen to illustrate different initial transverse offsets.

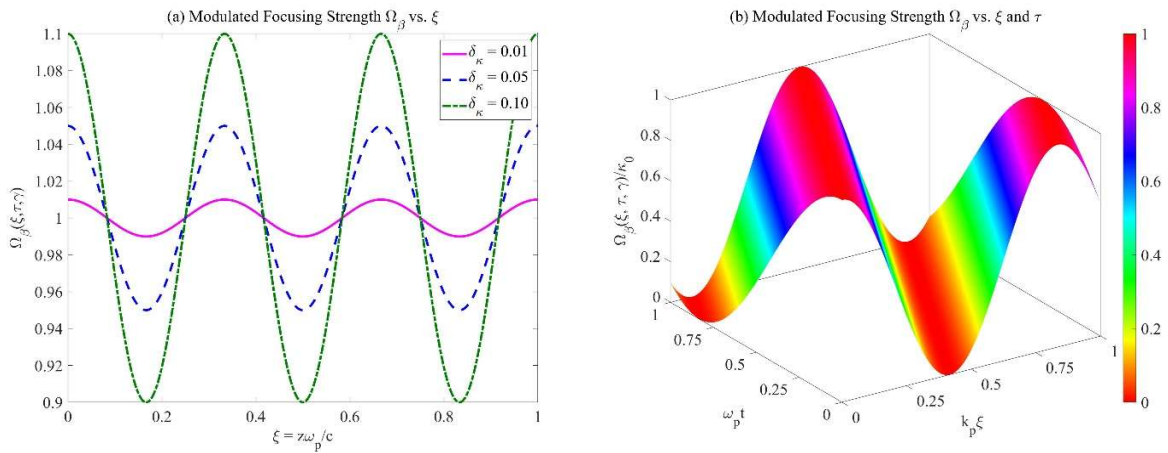

**Fig. 3.** (a) Plot of modulated focusing strength as a function of normalized longitudinal coordinate for different RF modulation amplitudes, (b) 3D surface plot of modulated focusing strength as a function of normalized of distance and time.

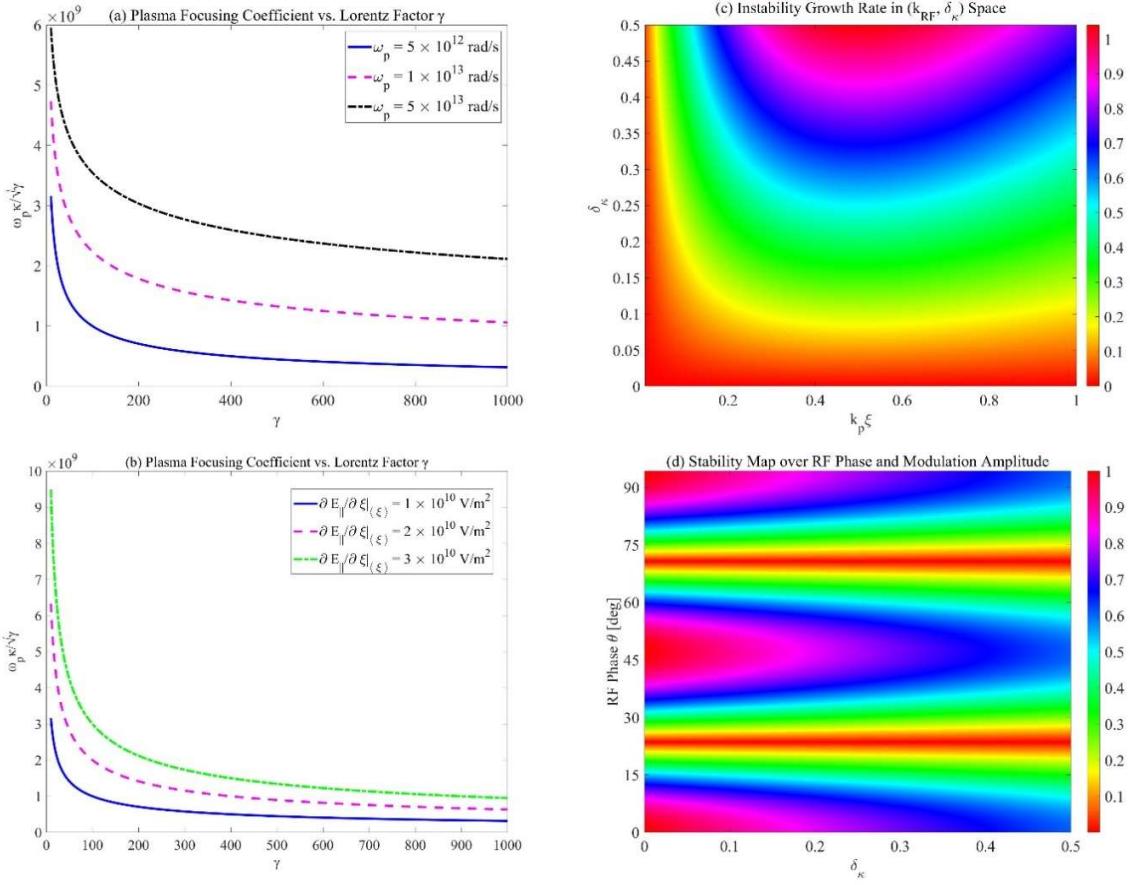

**Fig. 4.** Plot of plasma focusing coefficient as a function of Lorentz factor for different (a) plasma frequencies, and (b) longitudinal field gradients, (c) color map of instability growth rate versus of $\xi$ and $\delta_k$, (d) color map of stability parameter over RF phase and modulation amplitude.

Figures 5 and 6 indicate the normalised transverse force changes with plasma frequency, RF field strength, and phase. Figure 5a shows how the transverse force changes over time. At higher plasma frequencies, these changes happen more quickly and sharply. This means that the plasma wakefield interacts more strongly with the electron beam, leading the beam focus better and move in a more stable way. The changes in oscillation amplitudes show that frequency-dependent modulation affects the stability of the beam over time. Figure 5b shows that increasing the RF field strength increases the transverse force without changing its periodicity. This shows that the RF field is the main force behind electron motion and allows for precise changes to the trajectory. Figure 5c presents a 3D view where time and RF phase create wavy patterns, revealing phase-sensitive interactions that can either enhance or suppress the force. Normalized transverse force amplitude, in agreement with experimental observations that plasma frequency significantly increase wakefield strength [40, 41]. In the same way, Figure 6a demonstrates that increasing plasma frequency leads to larger and more rapidly oscillating points both in number and brightness thus causing the beam to be confined more efficiently but also acquire oscillatory energy. Figure 6b indicates that the rise of the RF field results in the increase in size and sharpness of the peaks thus confirming the RF field's role in the beam's transverse dynamics as the main driver. The change in amplitude of the oscillations does not affect their stable periodicity which is indicative of strong RF–plasma coupling that is vital for controlled acceleration. In Figure 6c, the 3D surface portrays the points of maximum and minimum smoothly shifting together with the phase of the RF thus signifying that phase modulation can be used to cancel out the oscillations effectively. These discoveries demonstrate the tactics that can be employed to enhance the performance of the accelerator and to generate ultra-relativistic electron beams having the desirable properties. Although the instantaneous Lorentz factor $\gamma(t)$ exhibits oscillations due to phase-dependent RF modulation, the cycle-averaged energy $\langle\gamma\rangle$ shows a net increasing trend; thus, "efficient energy transfer" refers to the enhanced cumulative energy gain per unit propagation length under optimal phase and frequency matching, rather than monotonic instantaneous acceleration.

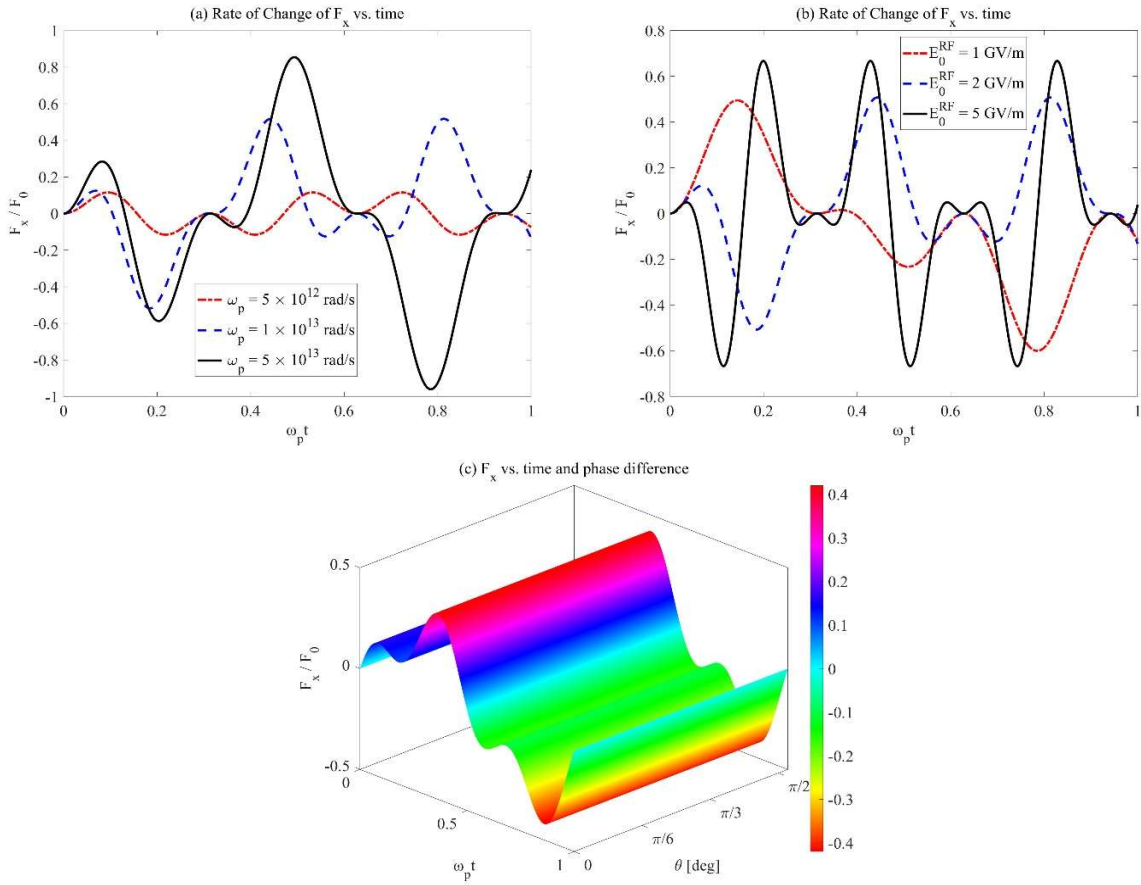

**Fig. 5.** Plot of normalized transverse force as a function of normalized time for different (a) plasma frequencies, and (b) RF electric field strengths, (c) 3D surface plot of normalized transverse force versus of normalized time and RF phase.

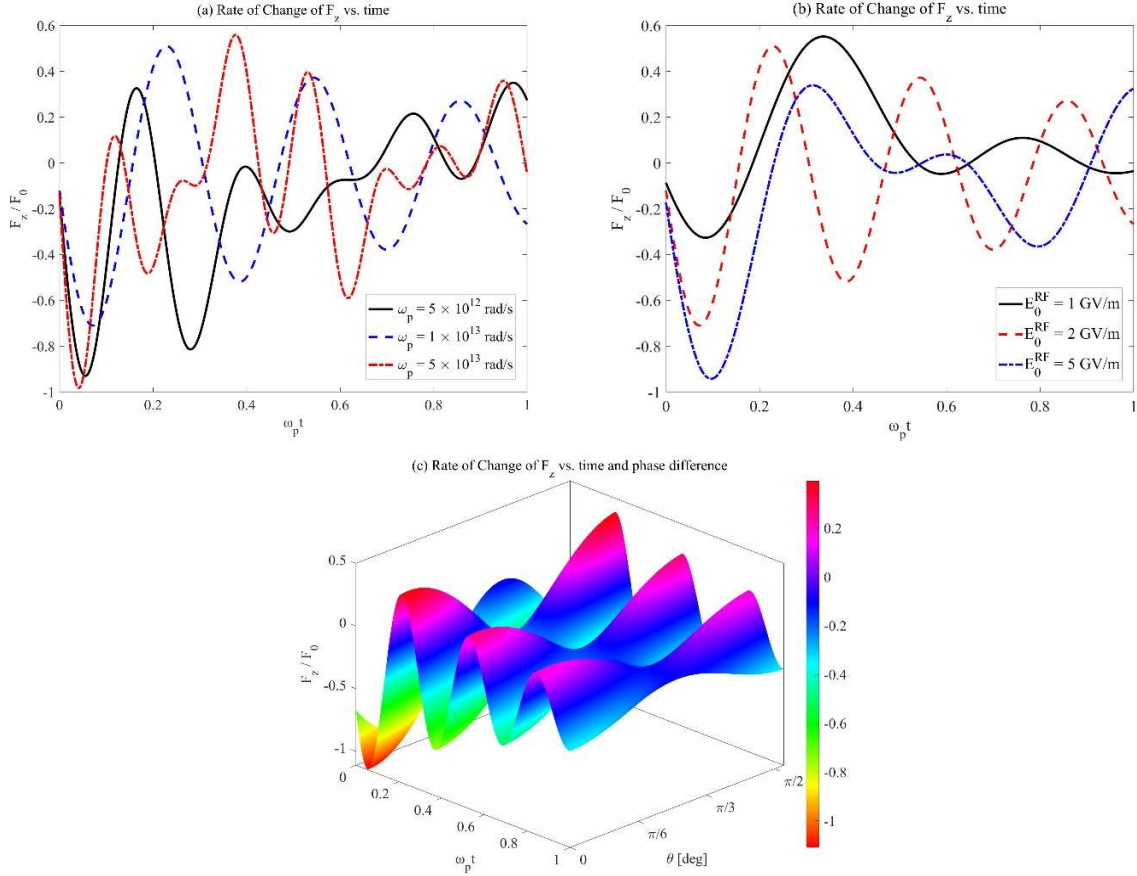

**Fig. 6.** Plot of the normalized longitudinal force in relation to normalized time for different (a) plasma frequencies and (b) RF electric field strengths, complemented by (c) a 3D surface depiction of the normalized longitudinal force with respect to normalized time and RF phase.

Understanding how the RF field parameters influence the electron beam's energy evolution is presented in Figure 7 and 8 for optimizing acceleration efficiency and beam stability in hybrid laser–plasma–RF systems. Figure 7a reveals a gradual increase in the average Lorentz factor over normalized time. The rate of growth becomes more pronounced as the RF electric field strength increases, suggesting that stronger RF fields enhance the electron beam's energy gain through more effective coupling with the plasma wakefield driving the longitudinal acceleration. It is important to note that the RF field is not meant to provide the main acceleration; instead, it is meant to cause controlled, phase-sensitive energy modulation. So, the small changes in $\gamma$ that we see here are a good thing because they keep the beam stable and the energy spread low while also allowing for precise control of the electron phase space. The small changes seen in the curves show that the energy evolution is stable but can be changed, which is important for keeping the beam quality at ultra-relativistic energies. Figure 7b indicates the impact of RF frequency on the temporal evolution of the Lorentz factor. As energy rises, higher frequencies add fast oscillations to the rising trend. This behaviour shows that for effective modulation of the acceleration process, the RF field and the plasma wake must be at the right frequency. This can be improved to make energy transfer more efficient. Figure 7c shows how the phase of the RF field affects the Lorentz factor profile. Different phase values create different patterns of peaks and troughs, showing a strong phase-dependent interaction that affects both the timing and the amount of energy gained. This provides a mechanism for fine-tuning the acceleration dynamics by carefully adjusting the RF phase. Figure 7d presents a smooth and colourful stability (or intensity) landscape as a function of the ratio of transverse electric field components and the RF phase. The observed gradients suggest that the polarization state of the emitted radiation, and the overall beam stability, are highly sensitive to these parameters. This mapping indicates potential optimal regions where maximum performance or minimal instability can be achieved. Furthermore, Figure 8a illustrates a rapid decline in the damped stability parameter as the longitudinal field gradient increases. This reduction is more notable at plasma frequencies suggesting that intensified wakefield gradients improve the suppression of instabilities. This suppression aids in stabilizing the beam by lessening transverse oscillations. Nevertheless, the impact slowly decreases at high gradients implying the existence of a saturation threshold. The clear difference between the curves for frequencies indicates how plasma frequency affects stability, and higher frequencies give you more control over beam coherence when focussing strongly. As the relativistic factor goes up, Figure 8b shows that the longitudinal divergence goes up as well. The rate at which this divergence grows changes with plasma frequency, which means that as electrons become more relativistic, the beam tends to spread out more in the longitudinal direction. This divergence is a little bigger at higher frequencies, which means that the energy is spread out more or the focussing isn't as good at high energies. The smooth upward trend shows that the evolution is predictable. This means that carefully adjusting the plasma parameters could lessen this effect and help keep the beam collimation in ultra-relativistic regimes.

Figures 9 and 10 show the betatron amplitude as a function of normalised time and distance for different parameters. Figures 9a–9d show how the damped phase changes over time. In the damped phase, Figure 9a shows oscillatory patterns where the amplitudes get bigger as the ratio of RF electric field components gets bigger. This means that stronger transverse field contributions make phase modulation stronger, which could cause bigger betatron oscillations and affect beam coherence. The different waveforms suggest that the response can be changed to make it more stable. Figure 9b shows that higher RF modulation amplitudes lead to stronger and faster phase oscillations. This shows that the RF field and the plasma wake are strongly linked, which can be used to control the rate of damping and lateral motion. But a lot of modulation can also cause instabilities that need to be handled with care. Figure 9c indicates that RF phase can affect by showing that some phase settings cause changes in amplitude and phase shifts. This phase-sensitive coupling could be used to match the RF field with the plasma wake to make acceleration more efficient or to stop unwanted oscillations. Figure 9d shows that the initial phase conditions are very important for long-term dynamics. When the starting values are higher, the oscillations last longer before they start to dampen. This shows how important it is to set the starting conditions carefully to get the stability characteristics you want. Figures 10a–10c look at how the betatron amplitude changes with different operating settings. Figure 10a shows that the normalised betatron amplitude is highest at intermediate wavenumber ratios. The plasma frequency changes the location and height of this peak, which means that resonance between the RF spatial frequency and the plasma wakefield makes transverse oscillations stronger. Higher frequencies make this resonance region wider, which could affect the stability of the beam. The amplitude goes down as the ratio goes up, which suggests a damping effect that could be used to control betatron motion. Figure 10b looks at the effect of the initial phase and shows that higher initial values lead to stronger and longer-lasting oscillations before a sharp drop. This shows how you can change the initial beam conditions to either improve or reduce transverse dynamics, depending on what you need. Figure 10c shows that when the RF modulation amplitude goes up, the betatron amplitude peak gets higher and wider. This effect increases radiation production, but it also makes things unstable, which must be controlled to keep the beam quality. The betatron amplitude is consistent with experimental findings indicating that plasma frequency, RF modulation, and initial phases substantially enhance wakefield strength [35, 38].

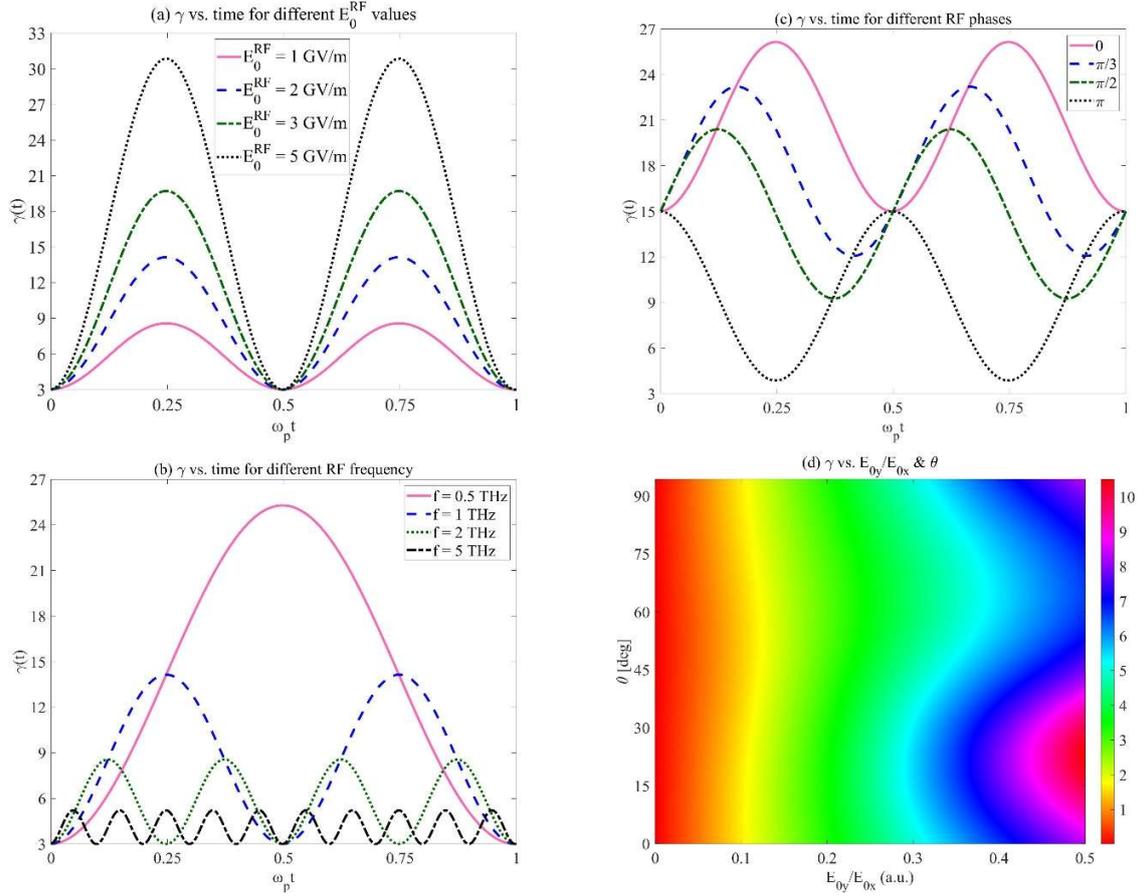

**Fig. 7.** Plot of average Lorentz factor as a function of normalized time for various (a) RF electric field strengths, (b) RF frequencies, (c) RF phases, and (d) color map of a parameter (e.g., energy or intensity) as a function of electric field ratio and RF phase.

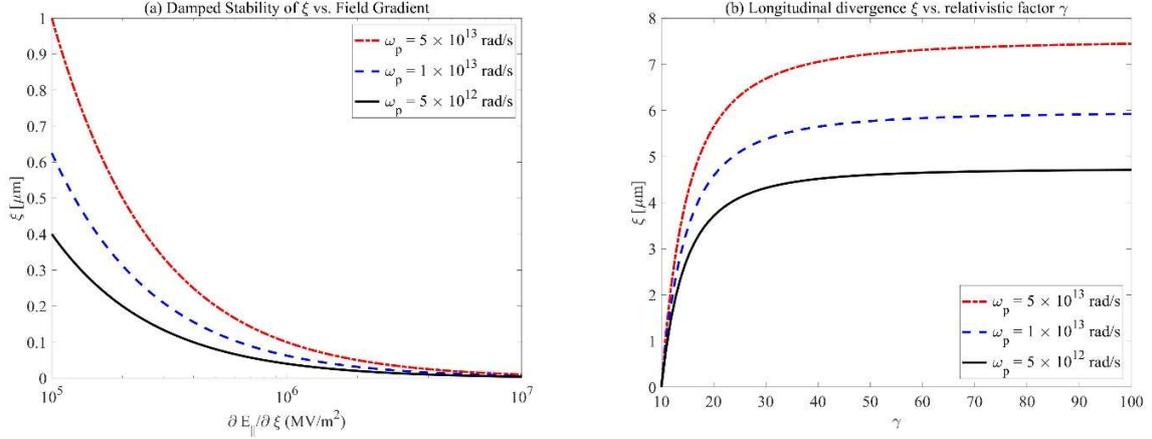

**Fig. 8.** (a) Plot of damped stability parameter as a function of longitudinal field gradient for different plasma frequencies, and (b) plot of longitudinal divergence versus relativistic factor for different plasma frequencies.

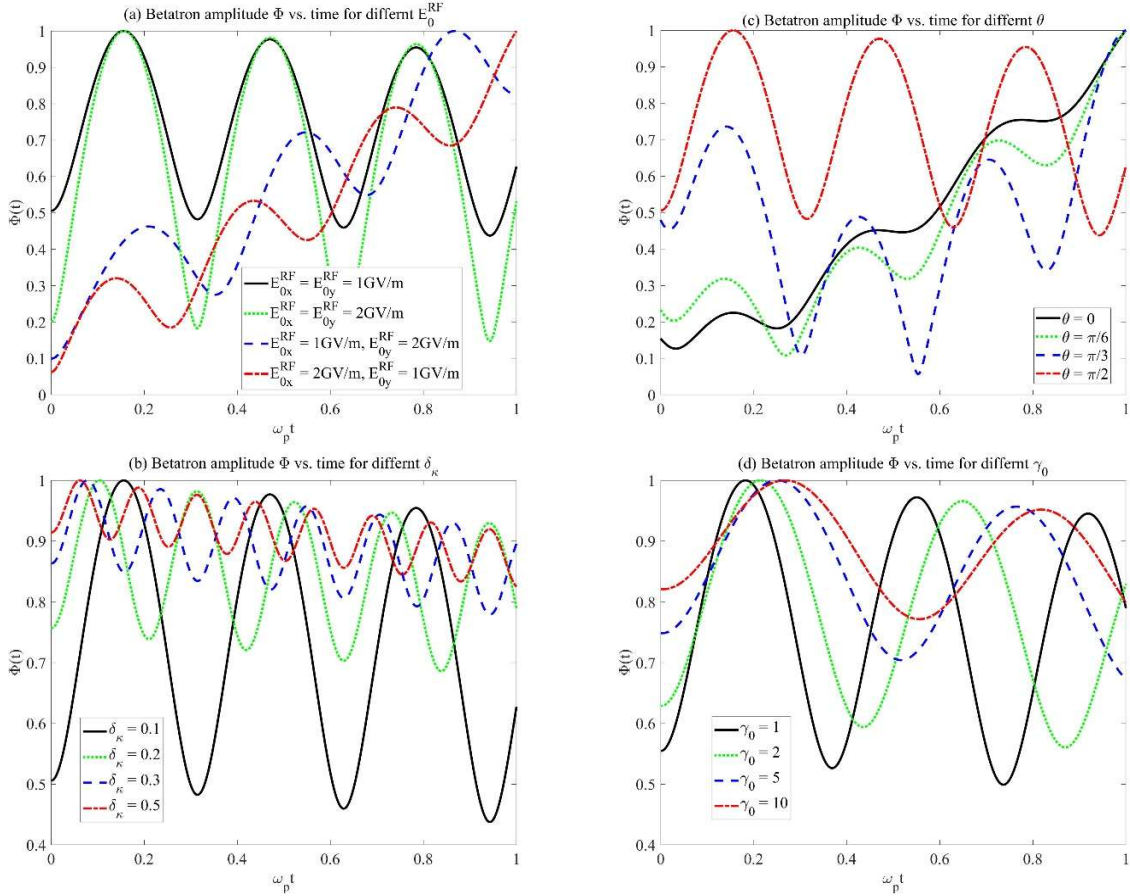

**Fig. 9.** Plot of betatron amplitude as a function of normalized time for various (a) RF electric field component ratios, (b) RF modulation amplitudes, (c) RF phases, and (d) initial phases.

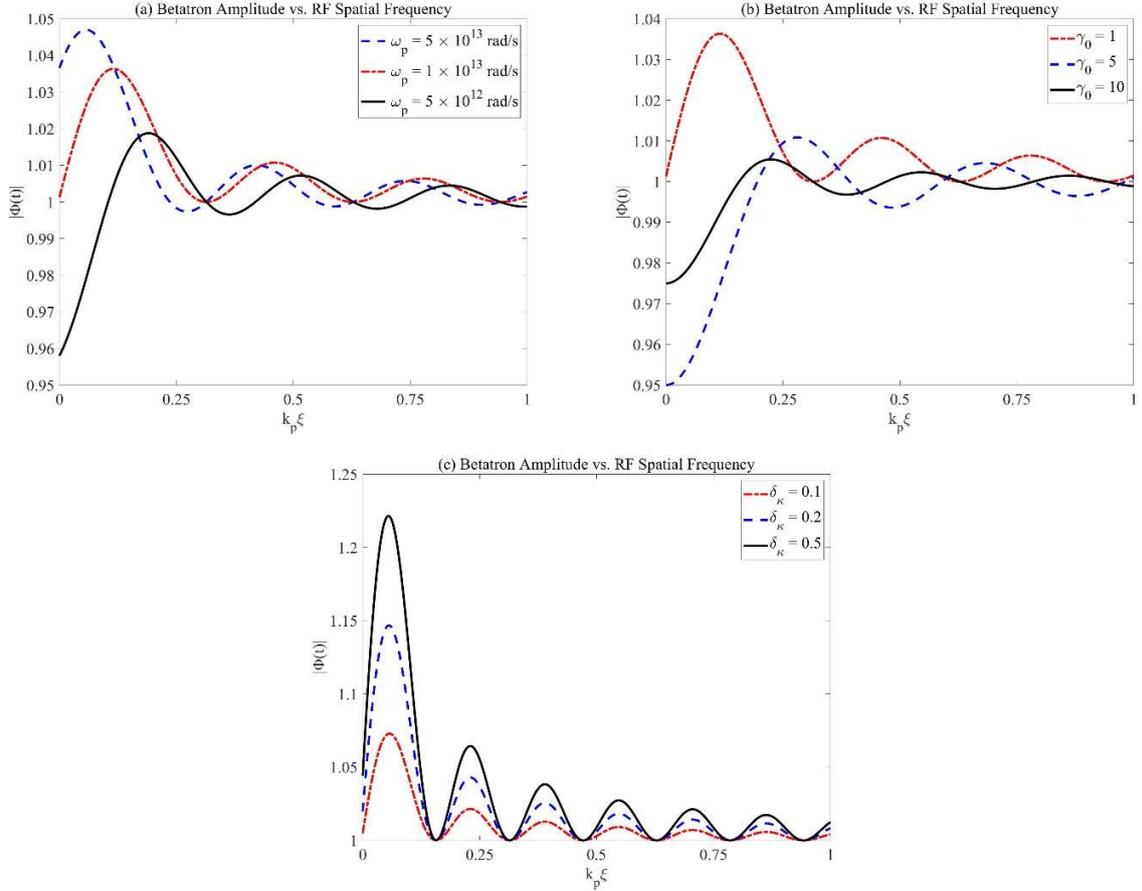

**Fig. 10.** Plot of betatron amplitude as a function of normalized distance for different (a) plasma frequencies, (b) initial phases, and (c) RF modulation amplitudes.

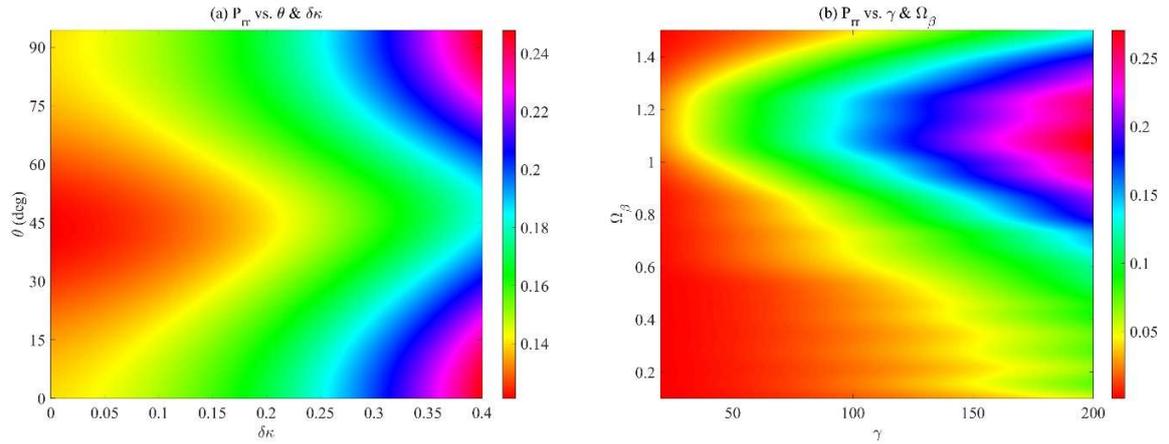

**Fig. 11.** Color map of normalized radiation reaction power ($P_{rr}$) as a function of (a) RF modulation amplitude and phase, (b) versus Lorentz factor and focusing strength.

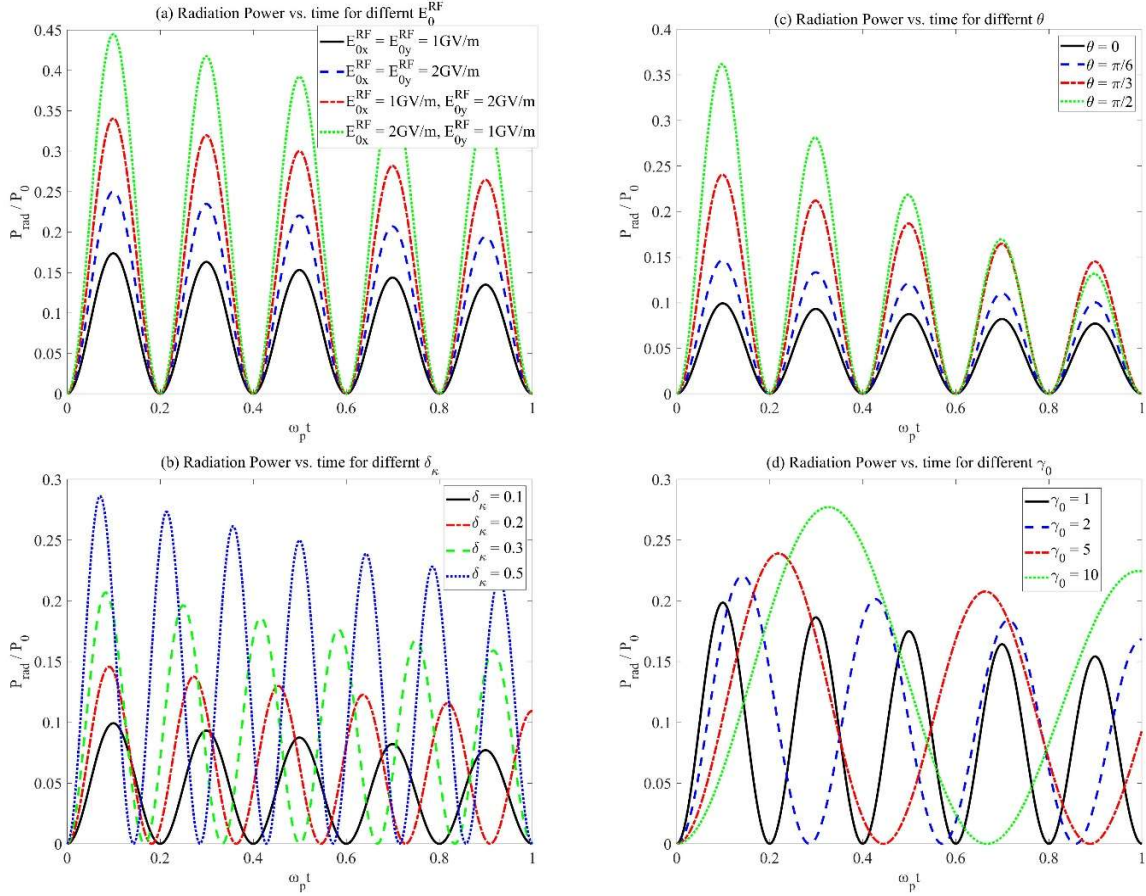

**Fig. 12.** Plot of radiation power as a function of normalized time for different (a) RF electric field strengths, (b) RF modulation amplitudes, (c) RF phases, and (d) initial relativistic factor.

Figures 11 and 12 show the radiation reaction power ($P_{rr}$) and radiation power for different parameters. Figure 11a indicates a symmetrical polarisation pattern with a central valley that gets deeper as the modulation amplitude and phase go up. According to the polarisation state is greatly affected by the interaction between RF modulation strength and phase alignment. The smooth gradient across the map suggests that it be changed at any time, which allows for customised radiation properties for specific experiments. At middle Lorentz factors and focussing strengths, Figure 11b shows a peak in polarisation. As relativistic effects increase, these peak shifts and widens, which means that higher betatron amplitudes make the emitted radiation more directional. When the parameters are very high or very low, the polarisation goes down, which shows that there may be limits to how long highly polarised emission can last. Also, Figure 12a shows that the normalised average longitudinal momentum keeps going up, and that higher RF field strengths make the trajectories curve up a little. This means that stronger RF fields help the plasma wakefield couple better, which speeds up the whole process. The RF field changes the longitudinal momentum, as shown in Figure 12b. Higher modulation amplitudes cause bigger oscillations, which gives you a way to adjust the balance between beam stability and acceleration efficiency. Figure 12c shows how RF phase affects things, with some phases giving bigger momentum gains or more regular oscillations. Aligning the phase with the plasma wake, then, makes it easier to move energy and cuts down on losses. The observed high-power emission aligns with experimental findings [42, 43].

Figure 13 indicates that the betatron energy in a hybrid laser-plasma-RF accelerator can change with the RF frequency and the phase difference between the two orthogonal RF field components. The three-dimensional plot shows that the external RF field and the plasma focussing fields work together in a complicated way. When the RF frequency changes, the betatron energy oscillates. There were both resonant and non-resonant states in this case. As the RF interacts harmonically with the natural betatron frequency, or frequency matches, it causes constructive interference. This creates stronger transverse oscillations, leading enhancement of the energy of the betatron. In contrast, when two waves are not in sync, destructive interference product, weakens the oscillations, which lowers the energy. The phase difference between the orthogonal RF components acts as a key control parameter. At specific phases, such as linear polarization, energy transfer becomes directional, favouring oscillations along one axis. With elliptical or circular polarization, energy spreads between both directions, introducing coupled transverse dynamics. This directly influences the polarization and angular distribution of the emitted betatron radiation. When the RF field and plasma wakefield interact synergistically, regions of maximum energy created, leading in strong focusing and efficient energy transfer. In contrast, minimum-energy regions indicate detuning and weaker coupling. By tuning the RF frequency and phase, betatron motion, the amplitude and polarization of emitted radiation can be precisely controlled. This causes generating ultra-relativistic electron beams for advanced light sources and compact free-electron lasers.

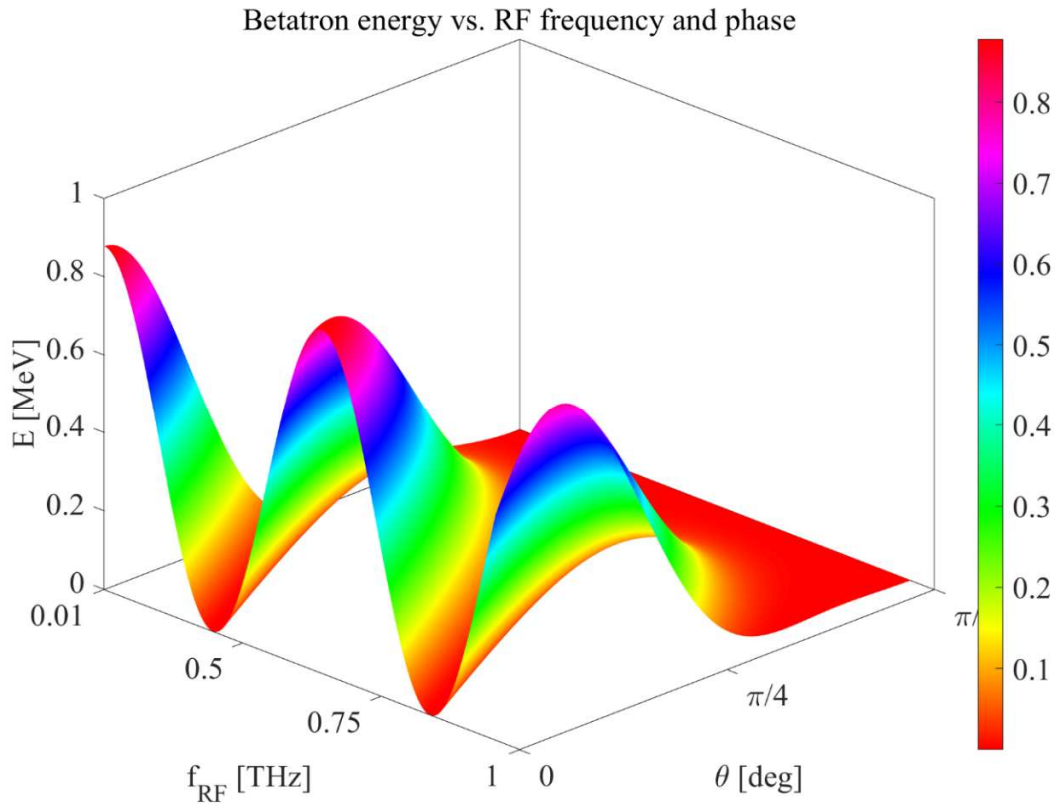

**Fig. 13.** Three-dimensional map of betatron energy as a function of RF frequency and phase difference.

## IV. Conclusions

This work presents a unified theoretical, computational, and physical framework for understanding and optimizing hybrid laser–plasma–RF accelerator architectures, with a particular focus on the coupled mechanisms that regulate transverse dynamics, radiation-reaction effects, and betatron polarization in ultra-relativistic electron beams. By combining analytical models of spatiotemporal wakefield structure, phase-dependent RF-induced transverse oscillations, and classical radiation reaction formulated through the Landau–Lifshitz equation, together with fully self-consistent three-dimensional particle-in-cell (3D-PIC) simulations using the EPOCH code, the multidimensional parameter space governing collective beam evolution is systematically dissected.

The results reveal that the interplay between plasma focusing forces and externally applied RF electromagnetic fields introduces a highly tunable effective lattice environment, in which the amplitude, frequency, and phase of the RF components act as direct control parameters for tailoring transverse focusing gradients and betatron oscillation amplitudes. In particular, fine adjustments of the RF amplitude and phase enable continuous manipulation of betatron polarization states, which exhibit pronounced resonant amplification as the RF frequency approaches the intrinsic betatron frequency of the plasma wakefield. The computed stability maps and fully resolved three-dimensional force landscapes further show that these resonant interactions not only allow controlled enhancement of transverse excursions but also suppress undesired parasitic oscillations through a combination of increased effective focusing strength and radiation-reaction-induced damping. This dual behavior provides a clear and physically transparent pathway for reducing normalized emittance and partially mitigating synchrotron-like radiation losses, even in regimes of high-energy gain. Here, the term "pathway" refers specifically to phase-synchronized RF modulation of the plasma focusing force, which enables deterministic control over the betatron amplitude, phase, and polarization. In the present hybrid scheme, this control is achieved by tuning a small set of experimentally accessible parameters, namely: (i) the RF frequency relative to the natural betatron frequency ($\omega_{RF} \approx \omega_\beta$), (ii) the normalized RF modulation amplitude α, typically in the range 0.01–0.1, (iii) the RF phase relative to the plasma wake phase, and (iv) the RF polarization state, controlled via a transverse phase offset.

Three-dimensional PIC simulations demonstrate that operation in resonant or near-resonant conditions, combined with appropriate selection of the RF phase to favour transverse damping rather than amplification, leads to systematic suppression of large-amplitude betatron oscillations. For representative ultra-relativistic electrons ($\gamma \gtrsim 200$), this controlled modulation reduces the projected normalized emittance by approximately 20–40% compared to the unmodulated plasma wakefield case over millimeter-scale propagation distances. In the same parameter regime, cumulative synchrotron-like radiation losses are reduced by approximately 10–20%, owing to the reduced transverse excursion and associated radiation emission. Moreover, the phase-sensitive nature of the hybrid accelerator dynamics is shown to govern the timing and persistence of transverse amplitude modulation over the acceleration length. Initial injection phase, RF carrier-envelope phase, and the relative ratio of longitudinal to transverse RF components collectively influence the coherence, stability, and long-term evolution of betatron motion. In the hybrid configuration, phase-synchronous RF coupling dynamically reshapes the effective longitudinal wakefield structure experienced by the electron bunch at fixed plasma density. Unlike conventional LWFA schemes, where the longitudinal gradient is primarily controlled by plasma density, the present hybrid RF–plasma system enables dynamic modulation of the effective accelerating gradient through RF phase, frequency, and amplitude, without altering the background plasma density. These results provide a quantitative and physically transparent strategy for tuning transverse beam quality in plasma-based accelerators through externally controlled RF–plasma coupling, without compromising the high-gradient acceleration provided by the plasma wake.

**Acknowledgements**
The authors acknowledge the use of the EPOCH particle-in-cell code, developed by the Warwick Plasma Physics Group.

**Disclosures**
The authors declare no conflicts of interest.

**Data availability**
The data that support the findings of this study are available from the corresponding author upon reasonable request.

**References**
[1] A. Curcio, G. Gatti, "Time-domain study of the synchrotron radiation emitted from electron beams in plasma focusing channels," Phys. Rev. E, 105, 025201, (2022), DOI: 10.1103/PhysRevE.105.025201.
[2] R. Rakowski, P. Zhang, K. Jensen, B. Kettle, T. Kawamoto, S. Banerjee, C. Fruhling, G. Golovin, D. Haden, M. S. Robinson, D. Umstadter, B. A. Shadwick, M. Fuchs, "Transverse oscillating bubble enhanced laser-driven betatron X-ray radiation generation," Sci. Rep., 2, 1, 10855, (2022), DOI: 10.1038/s41598-022-14748-z.
[3] A. M. de la Ossa, R. W. Assmann, M. Bussmann, S. Corde, J. P. C. Cabadağ, A. Debus, A. Döpp, A. F. Pousa, M. F. Gilljohann, T. Heinemann, B. Hidding, A. Irman, S. Karsch, O. Kononenko, T. Kurz, J. Osterhoff, R. Pausch, S. Schöbel, U. Schramm, "Hybrid LWFA-PWFA staging as a beam energy and brightness transformer: conceptual design and simulations," Philos Trans A Math Phys Eng Sci, 377, 2151, 20180175, (2019), DOI: 10.1098/rsta.2018.0175.
[4] R. D'Arcy, et. al, "plasma wakefield accelerator science for high-average-power applications," Philos Trans A Math Phys Eng Sci, 377, 2151, 20180392, (2019), DOI: 10.1098/rsta.2018.0392.
[5] M. Sedaghat, S. Barzegar, A. R. Niknam, "Quasi-phase-matched laser wakefield acceleration of electrons in an axially density-modulated plasma channel," Sci. Rep., 11, 1, 15207, (2021), DOI: 10.1038/s41598-021-94751-y.
[6] Z. Gong, F. Mackenroth, T. Wang, X. Q. Yan, T. Toncian, A. V. Arefiev, "Direct laser acceleration of electrons assisted by strong laser-driven azimuthal plasma magnetic fields," Phys. Rev. E, 102, 013206, (2020), DOI:10.1103/PhysRevE.102.013206.
[7] M. Shahmansouri,∗ B. Farokhi, and N. Khodabakhshi, "Modified Potential Around a Moving Test Charge in Strongly Coupled Dusty Plasma," Commun. Theor. Phys. 68 (2017) 111–116, (2017), DOI: 10.1088/0253-6102/68/1/111.
[8] A. A. Molavi Choobini , F. M. Aghamir, "Generation of THz waves through interaction of the wakefield of two-color laser pulses with magnetized plasma," J. Opt. Soc. Am. B, 42, 4, (2025), DOI: 10.1364/JOSAB.540636.
[9] Sh. Zhou, J. Hua, W. An, W. B. Mori, Ch. Joshi, J. Gao, W. Lu, "High Efficiency Uniform Wakefield Acceleration of a Positron Beam Using Stable Asymmetric Mode in a Hollow Channel Plasma," Phys. Rev. Lett., 127, 17, 174801, (2021), DOI: 10.1103/PhysRevLett.127.174801.
[10] A. A. Molavi Choobini , F. M. Aghamir, "Wakefield-induced THz wave generation in a hybrid dielectric-plasma cylindrical waveguide," Physics Letters A 559, 130910, (2025), DOI: 10.1016/j.physleta.2025.130910.

[11] K. G. Miller, J. R. Pierce, M. V. Ambat, J. L. Shaw, K. Weichman, W. B. Mori, D. H. Froula, J. P. Palastro, "Dephasing-less laser wakefield acceleration in the bubble regime," Sci. Rep., 13, 1, 21306, (2023), DOI: 10.1038/s41598-023-48249-4.

[12] A. F. Pousa, A. M. de la Ossa, R. W. Assmann, "Intrinsic energy spread and bunch length growth in plasma-based accelerators due to betatron motion," Sci. Rep., 9, 1, 17690, (2019), DOI: 10.1038/s41598-019-53887-8.

[13] R. Rakowski, P. Zhang, K. Jensen, B. Kettle, T. Kawamoto, S. Banerjee, C. Fruhling, G. Golovin, D. Haden, M. S. Robinson, D. Umstadter, B. A. Shadwick, M. Fuchs, "Transverse oscillating bubble enhanced laser-driven betatron X-ray radiation generation," Sci. Rep., 12, 1, 10855, (2022), DOI: 10.1038/s41598-022-14748-z.

[14] P. S. Miguel Claveria, E. Adli, L. D. Amorim, W. An, C. E. Clayton, S. Corde, S. Gessner, M. J. Hogan, C. Joshi, O. Kononenko, M. Litos, W. Lu, K. A. Marsh, W. B. Mori, B. O'Shea, G. Raj, D. Storey, N. Vafaei-Najafabadi, G. White, X. Xu, V. Yakimenko, "Betatron radiation and emittance growth in plasma wakefield accelerators," Philos Trans A Math Phys Eng Sci., 377, 2151, 20180173, (2019), DOI: 10.1098/rsta.2018.0173.

[15] V. Tomkus, V. Girdauskas, J. Dudutis, P. Gečys, V. Stankevič, G. Račiukaitis, I. G. González, D. Guénot, J. B. Svensson, A. Persson, O. Lundh, "Laser wakefield accelerated electron beams and betatron radiation from multijet gas targets," Sci. Rep., 10, 16807, (2020), DOI: 10.1038/s41598-020-73805-7.

[16] C. Joshi, S. Corde, W. B. Mori, "Perspectives on the generation of electron beams from plasma-based accelerators and their near- and long-term applications," Phys. Plasmas 27, 070602 (2020), DOI: 10.1063/5.0004039.

[17] R. Babjak, B. Martinez, M. Krus, M. Vranic, "Direct laser acceleration in varying plasma density profiles," New J. Phys. 26, 093002, (2024), DOI: 10.1088/1367-2630/ad7280.

[18] P. San Miguel Claveria, E. Adli, L. D. Amorim, W. An, C. E. Clayton, S. Corde, S. Gessner, M. J. Hogan, C. Joshi, O. Kononenko, M. Litos, W. Lu, K. A. Marsh, W. B. Mori, B. O'Shea, G. Raj, D. Storey, N. Vafaei-Najafabadi, G. White, X. Xu, V. Yakimenko, "Betatron radiation and emittance growth in plasma wakefield accelerators," Phil. Trans. R. Soc. A 377: 20180173, (2019), DOI: 10.1098/rsta.2018.0173.

[19] A. R. Maier, N. M. Delbos, T. Eichner, L. Hübner, S. Jalas, L. Jeppe, S. W. Jolly, M. Kirchen, V. Leroux, P. Messner, M. Schnepp, M. Trunk, P. A. Walker, Ch. Werle, P. Winkler, "Decoding Sources of Energy Variability in a Laser-Plasma Accelerator," Physical Review X, 10, 031039 (2020), DOI: 10.1103/PhysRevX.10.031039.

[20] A. Köhler, J. P. Couperus, O. Zarini, A. Jochmann, A. Irman, U. Schramm, "Single-shot betatron source size measurement from a laser-wakefield accelerator," Nuclear Instruments & Methods in Physics Research A, (2016), DOI: 10.1016/j.nima.2016.02.031i.

[21] H. G. Rinderknecht, T. Wang, A. L. Garcia, G. Bruhaug, M. S. Wei, H. J. Quevedo, T. Ditmire, J. Williams, A. Haid, D. Doria, K. M. Spohr, T. Toncian, A. Arefiev, "Relativistically transparent magnetic filaments: scaling laws, initial results and prospects for strong-field QED studies," New J. Phys. 23, 095009, (2021), DOI: 10.1088/1367-2630/ac22e7.

[22] J. Cikhardt, M. Gyrdymov, S. Zähter, P. Tavana, M. M. Günther, N. Bukharskii, N. Borisenko, J. Jacoby, X. F. Shen, A. Pukhov, N. E. Andreev, O. N. Rosmej, "Characterization of bright betatron radiation generated by direct laser acceleration of electrons in plasma of near critical density," Matter Radiat. Extremes 9, 027201 (2024), DOI: 10.1063/5.0181119.

[23] O. N. Rosmej, X. F. Shen, A. Pukhov, L. Antonelli, F. Barbato, M. Gyrdymov, M. M. G¨unther, S. Z¨ahter, V. S. Popov, N. G. Borisenko, N. E. Andreev, "Bright betatron radiation from direct-laseraccelerated electrons at moderate relativistic laser intensity," Matter Radiat. Extremes, 6, 048401 (2021), DOI: 10.1063/5.0042315.

[24] M. Yadav, C. Hansel, B. Naranjo, G. Andonian, P. Manwani, Ö. Apsimon, C. P. Welsch, J. Rosenzweig, "Modeling betatron radiation using particle-in-cell codes for plasma wakefield accelerator diagnostics," Physical Review Accelerators and Beams, 28, 072801, (2025), DOI: 10.1103/ykrl-45h8.

[25] A. Curcio, A. Cianchi, G. Costa, A. D. Dotto, F. Demurtas, M. Ferrario, M. D. R. Frías, M. Galletti, J. A. Pérez-Hernández, G. Gatti, "Reconstruction of lateral coherence and 2D emittance in plasma betatron X-ray sources," Sci. Rep., 14, 1719, (2024), DOI: 10.1038/s41598-024-52231-z.

[26] J. P. Farmer, G. Zevi Della Porta, "Wakefield regeneration in a plasma accelerator," Phys. Rev. Research 7, L012055, (2025), DOI: 10.1103/PhysRevResearch.7.L012055.

[27] P. Winkler, M. Trunk, L. Hübner, A. Martinez de la Ossa, S. Jalas, M. Kirchen, I. Agapov, S. A. Antipov, R. Brinkmann, T. Eichner, A. Ferran Pousa, T. Hülsenbusch, G. Palmer, M. Schnepp, K. Schubert, M. Thévenet, P. A. Walker, C. Werle, W. P. Leemans, A. R. Maie, "Active energy compression of a laser-plasma electron beam," Nature 640, 907–910 (2025), DOI: 10.1038/s41586-025-08772-y.

[28] Th. P. Wangler, "RF Linear Accelerators," Wiley-VCH, 2nd, Completely Revised and Enlarged, (2008), ISBN-10: 3527406808.

[29] E. Esarey, C. B. Schroeder, W. P. Leemans, "Physics of laser-driven plasma-based electron accelerators," Rev. Mod. Phys. 81, 1229, (2009), DOI: 10.1103/RevModPhys.81.1229.

[30] T D Arber, et., al, "Contemporary particle-in-cell approach to laser-plasma modelling," Plasma Phys. Control. Fusion 57 113001, (2015), DOI: 10.1088/0741-3335/57/11/113001.

[31] Marija Vranic, et., al, "Quantum radiation reaction in head-on laser-electron beam interaction," New J. Phys. 18 073035, (2016), DOI: 10.1088/1367-2630/18/7/073035.

[32] Ridgers CP, Blackburn TG, et., al, "Signatures of quantum effects on radiation reaction in laser–electron-beam collisions," Journal of Plasma Physics., 83 (5), 715830502, (2017), DOI:10.1017/S0022377817000642.

[33] M. T. Hibberd, A. L. Hearly, D. S. Lake, et., al, "Acceleration of relativistic beams using laser-generated terahertz pulses," Nat. Photonics 14, 755–759 (2020), DOI: 10.1038/s41566-020-0674-1.

[34] Y. X. Wang, et., al, "High-efficiency and frequency-controllable terahertz pulses driven by two-color relativistic laser pulses," Phys. Plasmas 31, 033111, (2024), DOI: 10.1063/5.0191984.

[35] Michaela Kozlova, et., al, "Hard X Rays from Laser-Wakefield Accelerators in Density Tailored Plasmas," Physical Review X, 10, 011061 (2020), DOI: 10.1103/PhysRevX.10.011061.

[36] Rafal Rakowski, et., al, "Transverse oscillating bubble enhanced laser-driven betatron X-ray radiation generation," Scientific Reports, 12, 10855, (2022), DOI: 10.1038/s41598-022-14748-z.

[37] Mengyuan Chu, et., al, "Controlled Betatron radiation from high-charge electron beams in multiple plasma channels," Optics Express, 33, 10, (2025), DOI: 10.1364/OE.557855.

[38] S. Y. Kim, et., al, "Witness electron beam injection using an active plasma lens for a proton beam-driven plasma wakefield accelerator," Physical Review Accelerators and Beams, 24, 121304 (2021), DOI: 10.1103/PhysRevAccelBeams.24.121304.

[39] Eitan Y. Levine, et., al, "Direct visualization of shock front induced nonlinear laser wakefield dynamics," Physical Research, 7, L012041, (2025), DOI: 10.1103/PhysRevResearch.7.L012041.

[40] Linbo Liang, et., al, "Simulation study of betatron radiation in AWAKE Run 2 experiment," arXiv:2204.13199v1.

[41] S. Mishra, et., al, "Enhanced betatron x-ray emission in a laser wakefield accelerator and wiggler due to collective oscillations of electrons," Physical Review Accelerators and Beams, 25, 090703 (2022), DOI: 10.1103/PhysRevAccelBeams.25.090703.

[42] Alessandro Curcio, et., al, "Performance Study on a Soft X-ray Betatron Radiation Source Realized in the Self-Injection Regime of Laser-Plasma Wakefield Acceleration," Appl. Sci., 12, 12471, (2022), DOI: 10.3390/app122312471.

[43] Srimanta Maity, et., al, "Coupling and Acceleration of Externally Injected Electron Beams in Laser-Driven Plasma Wakefields," arXiv:2502.10160.